\documentclass[aip,amsmath,amssymb,reprint,groupedaddress,floatfix]{revtex4-1}

\usepackage{graphicx}
\usepackage{floatrow}
\usepackage[caption=false,font={large}]{subfig}
\usepackage{bm}

\usepackage{amsmath}
\usepackage{amssymb}
\usepackage{color}
\usepackage{times}
\usepackage{float}
\usepackage[sort&compress]{natbib}

\usepackage{hyperref}
\hypersetup{
  colorlinks=true,
citecolor=black,
linkcolor=black,
urlcolor=black
  }

\usepackage[T1]{fontenc}
\usepackage[utf8]{inputenc}

\newcommand{\xm}{x_{\mathrm{m}}}

\begin{document}

\title{Resetting induced multimodality}

\author{Przemys{\l}aw Pogorzelec}
\email{przemyslaw.pogorzelec@doctoral.uj.edu.pl}
\affiliation{Doctoral School of Exact and Natural Sciences, Faculty of Physics, Astronomy and Applied Computer Science,
Jagiellonian University, \L{}ojasiewicza 11, 30-348 Krak\'ow, Poland}

\author{Bart{\l}omiej Dybiec}
\email{bartlomiej.dybiec@uj.edu.pl}

\affiliation{Institute of Theoretical Physics,
and Mark Kac Center for Complex Systems Research,
Faculty of Physics, Astronomy and Applied Computer Science,
Jagiellonian University, \L{}ojasiewicza 11, 30-348 Krak\'ow, Poland}


\date{\today}

\begin{abstract}
Properties of stochastic systems are defined by the noise type and deterministic forces acting on the system.
In out-of-equilibrium setups, e.g., for motions under action of L\'evy noises, the existence of the stationary state is not only determined by the potential but also by the noise.
Potential wells need to be steeper than parabolic in order to assure existence of stationary states.
The existence of stationary states, in sub-harmonic potential wells, can be restored by stochastic resetting, which is the protocol of starting over at random times.
Herein we demonstrate that the combined action of L\'evy noise and Poissonian stochastic resetting can result in the phase transition between non-equilibrium stationary states of various multimodality in the overdamped system in super-harmonic potentials.
Fine-tuned resetting rates can increase the modality of stationary states, while for high resetting rates the multimodality is destroyed as the stochastic resetting limits the spread of particles.
\end{abstract}

\pacs{02.70.Tt,
 05.10.Ln, 
 05.40.Fb, 
 05.10.Gg, 
  02.50.-r, 
  }

%
\maketitle

\setlength{\tabcolsep}{0pt}

\textbf{Under action of the Gaussian white noise the form of a stationary state reflects the shape of the potential, because stationary states are of the Boltzmann-Gibbs type.
Consequently, the number of modes in the stationary state is the same as the number of minima of the potential.
The situation is very different in the non-equilibrium regime, e.g., under action of the L\'evy noise.
For instance, action of the Cauchy noise can induce a bimodal stationary state in a single well potential.
Here, we demonstrate that stochastic resetting can be used as a measure of controlling the modality of non-equilibrium stationary states.
In particular we show that simultaneous action of stochastic resetting and L\'evy noise can result in a further increase in the number of modal values.
}

%
%
\section{Introduction}

The problem of stationary states in stochastic systems has been studied for a very long time both in the regime of full (underdamped) \cite{risken1996fokker,magdziarz2007c} and overdamped \cite{gardiner2009,chechkin2002,dubkov2007} dynamics.
For motions in power-law potentials
$   V(x) \propto |x|^c,
	\label{eq:powerPotential}
$
with $c>0$, under the action of the Gaussian white noise, the stationary states always exist.
The situation becomes more subtle when the Gaussian white noise is replaced with the L\'evy noise, which is frequently used to describe out-of-equilibrium systems.
The problem of the existence of stationary states in overdamped systems driven by L\'evy noise is well understood and widely explored \cite{jespersen1999,chechkin2002,chechkin2003,chechkin2004}.
Stationary states exist for power-law potentials with $c>2-\alpha$, see Ref.~\onlinecite{dybiec2010d}, where $\alpha$ is the stability index determining power-law asymptotics of $\alpha$-stable densities.
They are not of the Boltzmann-Gibbs type \cite{eliazar2003}, albeit for $c=2$ (harmonic potential)  are given by the $\alpha$-stable density determining the noise type, i.e., the $\alpha$-stable density with the same value of the stability index as the driving noise.
Remarkably, for $c>2$ stationary states not only exist but are bimodal \cite{chechkin2003,chechkin2004}.
In the underdamped case, the problem can be more complex as the nonlinear friction tames L\'evy flights and weakens the condition on the minimal value of the exponent $c$, see Ref.~\onlinecite{capala2020nonlinear}.

Another process that can influence the problem of existence of stationary states is stochastic resetting \cite{evans2011diffusion,evans2020stochastic,gupta2022stochastic}.
Typically, stationary states do not exist if a particle, despite the restoring forces, can escape to infinity with non-negligible probability.
Stochastic resetting starts the motion anew, efficiently eliminating long excursions and limiting the spread of particles.
This in turn allows for the emergence of non-equilibrium stationary states in diverse situations, including free motion \cite{evans2011diffusion,nagar2016diffusion}, L\'evy flights \cite{stanislavsky2021optimal}, continuous time random walks\cite{mendez2021ctrw} or even motion in inverted potentials \cite{pal2015diffusion}.

The stochastic resetting\cite{evans2011diffusion,evans2020stochastic,gupta2022stochastic} assumes that the motion is started anew at random times.
The natural approach is to assume that restarts are triggered temporally, i.e., times of starting over are independent of state of the system, e.g., position.
For example, resets can be performed periodically (sharp resetting)\cite{pal2017first} or
at random exponentially (Poissonian resetting)\cite{evans2011diffusion}, or power-law distributed  \cite{nagar2016diffusion} time intervals.
In addition to temporal resetting, starting anew can be also spatially induced \cite{dahlenburg2021stochastic}.
Finally, resetting does not need to be hard but it can be soft \cite{xu2022stochastic} in the sense that instead of bringing a particle back to a starting position an additional deterministic force capable of moving a particle towards a given point is turned on.

Stochastic resetting is tightly connected with search strategies \cite{reynolds2009,viswanathan2011physics,palyulin2014levy}, which in turn are related to the first passage problems \cite{redner2001} because in the search strategy one typically optimizes the time needed to find a target.
The stochastic resetting is not only capable of making the mean first passage time (MFPT) finite, in setups where due to long excursions to points distant from the target\cite{kusmierz2014firstorder,mendez2021ctrw} it diverges, but can also optimize\cite{reuveni2016optimal,pal2017first} already finite MFPT.
More precisely, in situations when the coefficient of variation
(the ratio between the standard deviation of the first passage times and the MFPT in the absence of
stochastic resetting)
is greater than unity\cite{reuveni2016optimal,pal2017first}, the stochastic resetting is beneficial and it is possible to find the optimal resetting rate resulting in the minimal value of the completion time and consequently optimize the search process.

Here, we are interested in the interplay between stochastic resetting and action of L\'evy noises with special attention to the emergence of multimodal stationary states.
The stochastic resetting can produce a cusp \cite{pal2015diffusion,stanislavsky2021optimal} at the position from which the motion is restarted.
Therefore, if the resetting to a single, fixed, position is replaced with the resetting to a discrete set of points, the stationary density can naturally become multimodal, because the stationary density is the superposition of stationary states corresponding to various restarting points \cite{evans2011diffusion-jpa,evans2020stochastic}.
The composition of the densities arises due to the linearity of the diffusion equation.
Here, we are assuming that the motion is driven by the L\'evy noise which typically produces multimodal stationary states \cite{capala2019multimodal}.
Therefore, we are exploring the possibility of further increase in the number of modal values thanks to the stochastic resetting to a single, fixed, point only.

%
%
\section{Model\label{sec:model}}

Within the current study, we explore the multimodality of stationary states under the combined action of stochastic resetting \cite{evans2011diffusion,christou2015diffusion,eule2016non,evans2020stochastic} and $\alpha$-stable noise \cite{bardou1990levystatistics,bassingthwaighte1994,dubkov2008}.
The particle motion is described by the following overdamped Langevin equation
\begin{equation}
	\frac{dx}{dt} = -V'(x ) + \xi_\alpha(t) = -x^3 + \xi_\alpha(t),
	\label{eq:langevin}
\end{equation}
where  $\xi_\alpha(t)$ stands for the $\alpha$-stable noise and $V(x)=x^4/4$ is the quartic potential producing the deterministic restoring force $-x^3$.

The $\alpha$-stable noise is a generalization of the Gaussian white noise to the non-equilibrium realms \cite{janicki1994}.
Within the current research, we restrict ourselves to symmetric $\alpha$-stable noise only, which is the formal time derivative of symmetric $\alpha$-stable process $L(t)$, see Refs.~\onlinecite{janicki1994,dubkov2008}.
Increments $\Delta L=L(t+\Delta t)-L(t)$ of the $\alpha$-stable process are independent and identically distributed according to an $\alpha$-stable density.
Symmetric $\alpha$-stable densities are unimodal probability densities, with the characteristic function \cite{samorodnitsky1994,janicki1994}
\begin{equation}
\varphi(k)  = \exp\left[ - \sigma^{\alpha}\vert k\vert^{\alpha} \right].
	\label{eq:levycf}
\end{equation}
The stability index $\alpha$ ($0<\alpha \leqslant 2$) determines the tail of the distribution, which for $\alpha<2$ is of power-law type $p(x) \propto |x|^{-(\alpha+1)}$.
The scale parameter $\sigma$ ($\sigma>0$) controls the width of the distribution, typically defined by an interquantile width or by fractional moments $\langle |x|^\kappa \rangle$ ($0<\kappa<\alpha$), because the variance of $\alpha$-stable variables diverges \cite{samorodnitsky1994,weron1995} for $\alpha<2$.

The Langevin equation is approximated with the (stochastic) Euler--Maruyama method \cite{higham2001algorithmic,mannella2002}
\begin{equation}
x(t+\Delta t) = x(t) -x^3(t) \Delta t + \xi_\alpha^t \Delta t^{1/\alpha}.
	\label{eq:integration}
\end{equation}
In Eq.~(\ref{eq:integration}) $\xi_\alpha^t$ represents the sequence of independent identically distributed $\alpha$-stable random variables \cite{chambers1976,weron1995,weron1996}, see Eq.~(\ref{eq:levycf}).

For the L\'evy noise with $\alpha=1$ (Cauchy noise), the non-equilibrium stationary density for $V(x)=x^4/4$ reads \cite{chechkin2002,chechkin2003,chechkin2004,chechkin2006,chechkin2008introduction}
\begin{equation}
	p_{\alpha=1}(x)=\frac{1}{\pi\sigma^{\frac{1}{3}}\left[\left(\frac{x}{\sqrt[3]{\sigma}}\right)^4-\left(\frac{x}{\sqrt[3]{\sigma}}\right)^2+1\right]}
	\label{eq:stationary-n4}.
\end{equation}
The stationary state~(\ref{eq:stationary-n4}) is a symmetric bimodal distribution with modes at $x=\pm \sigma^{\frac{1}{3}}/\sqrt{2}$ and
the power-law asymptotics $p(|x|) \propto |x|^{-4}$.
The recorded bimodality in Eq.~(\ref{eq:stationary-n4}) is related to the general property of the  stationary states in superharmonic potentials under action of L\'evy noise~\cite{chechkin2002,chechkin2003,dubkov2007,capala2019multimodal}, i.e., their bimodality.

\begin{figure}[!h]
	\centering
   	  \includegraphics[angle=0,width=0.95\columnwidth]{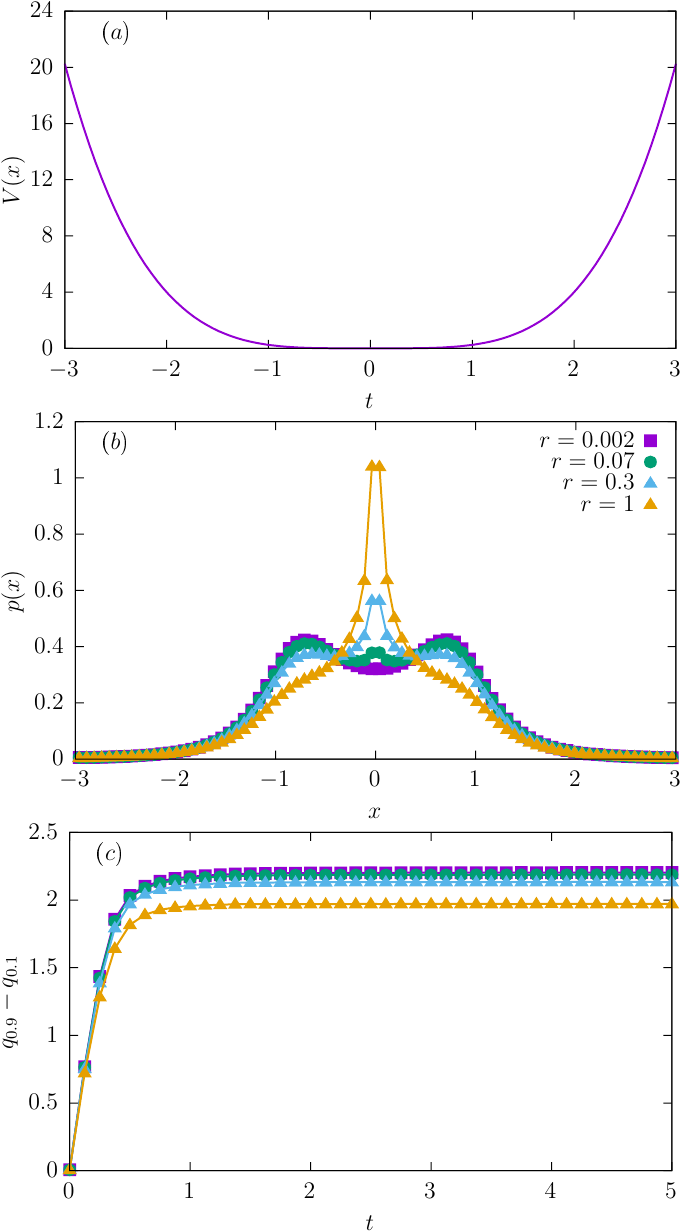}\\
   	  \caption{Quartic $V(x)=x^4/4$ potential (top panel -- ($a$)), non-equilibrium stationary states for the motion in the quartic potential (middle panel -- ($b$)) and interquantile distances (bottom panel -- ($c$)) under combined action of the stochastic resetting and Cauchy noise.
   	  Various points correspond to various resetting rates $r$, $r\in\{0.002,0.07,0.3,1\}$.}
	\label{fig:st-x4}
\end{figure}

The time evolution of the full probability density associated with Eq.~(\ref{eq:langevin}) is described by the fractional Smoluchowski-Fokker-Planck equation  \cite{risken1996fokker,jespersen1999,kilbas2006,podlubny1999,yanovsky2000,schertzer2001}
\begin{equation}
 \frac{\partial p }{\partial t}=  \frac{\partial}{\partial x}  \left[   V'(x) p \right] +  \sigma^\alpha \frac{\partial^\alpha p}{\partial |x|^\alpha},
 \label{eq:ffpe}
\end{equation}
\normalsize
where $p=p(x,t|x_0,t_0)$.
The fractional  Riesz-Weil derivative  \cite{podlubny1999,samko1993}  is understood in the sense of the Fourier transform
\begin{equation}
\mathcal{F}_k\left( \frac{\partial^\alpha f(x) }{\partial |x|^\alpha} \right)=-|k|^\alpha \mathcal{F}_k(f(x)).
\end{equation}
In further considerations, we use Cauchy noise ($\alpha$-stable noise with $\alpha=1$). Additionally, we set the scale parameter $\sigma$ to unity, i.e., $\sigma=1$.

The motion described by Eq.~(\ref{eq:langevin}) is affected by the Poissonian stochastic resetting \cite{evans2020stochastic}.
At random time instants, the particle position $x(t)$ is reset to $x_0$ and the duration of interresetting time intervals $\tau$ is exponentially distributed, i.e., $p(\tau) = r\exp(-r \tau )$, where $r>0$ is the resetting rate.
The mean interresetting time is equal to $\langle \tau \rangle = \frac{1}{r}$.
The stochastic resetting typically leads to the emergence of the cusp (mode) at $x_0$ \cite{evans2011diffusion-jpa,pal2015diffusion,evans2020stochastic,stanislavsky2021optimal}.
The emergence of the cusp (mode), combined with the fact that a stationary state for resetting to multiple points is the composition of stationary states corresponding to each restarting point, see Appendix~\ref{sec:res-app}, suggests an easy method of producing multimodal stationary states.
Instead of restarting the motion from a fixed point, the motion needs to be started over from randomly sampled points from a given discrete set of points, see Appendix~\ref{sec:res-app}.
However, here we explore a less obvious approach, which is limited to restarting from a fixed point only, i.e., minimum of the potential.
Putting it differently, we study competition between diffusive spread (exploration of space) of particles and resetting (reintroduction of particles to a given point), which limits spread of particles.

The combined action of the non-equilibrium noise and stochastic resetting is studied numerically.
Obtained numerical results, see Sec.~\ref{sec:results}, have been constructed by the approximation~(\ref{eq:integration}) with $\Delta t=10^{-3}$ and averaged over $10^7$ realizations constructed by the Euler-Maruyama method \cite{higham2001algorithmic,mannella2002,janicki1994,janicki1994b}.

%
%
\section{Results\label{sec:results}}

\begin{figure}[!h]
	\centering
\includegraphics[angle=0,width=0.95\columnwidth]{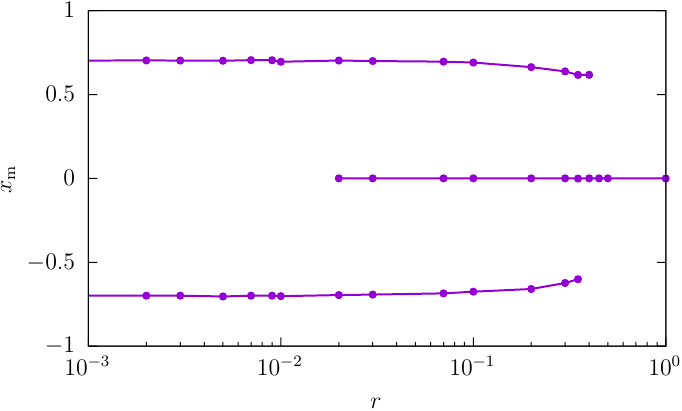}\\
   	  \caption{The phase diagram showing the location of modal values $\xm$ of non-equilibrium stationary states for the motion in the quartic potential under combined action of the stochastic resetting and Cauchy noise as a function of the resetting rate $r$.
   	  }
	\label{fig:pd-x4}
\end{figure}

The model under study is built by three components: stochastic force (noise), deterministic force and stochastic resetting.
Noise is responsible for the spread of particles.
Since the noise is of the $\alpha$-stable type it can easily produce long jumps which result in  facilitated exploration of the space.
The deterministic restoring force not only assures existence of stationary states, but also moves particles towards the origin.
Finally, the stochastic resetting introduces a source of particles  at $x_0$ (origin) and a sink at every other location.
The source of particles is capable of building a mode at the origin, while the sink prevents exploration of the space at longer length-scales.
Overall, the observed stationary state is the outcome of the competition between model components: especially between noise induced exploration at longer scales and restraining action of the stochastic resetting.

Despite the described straightforward method of producing multimodal stationary states by restarting from multiple points, here, we are exploring the standard, Poissonian, resetting to a fixed point.
Such a standard resetting combined with the action of L\'evy noise is capable of affecting the multimodality of stationary states.
First, we explore the motion in a quartic $V(x)=x^4/4$ potential.
In the absence of resetting, under action of the Cauchy noise ($\alpha$-stable noise with $\alpha=1$), the stationary state is given by Eq.~(\ref{eq:stationary-n4}).

Fig.~\ref{fig:st-x4} presents the quartic potential (top panel -- ($a$)) and sample stationary states under stochastic resetting (middle panel -- ($b$)).
In order to verify that the stationary states have been reached we have checked that interquantile distances stagnate at constant values, see Fig.~\ref{fig:st-x4}(c).
For a very low $r$ the stationary state is practically identical to the one without resetting, see Eq.~(\ref{eq:stationary-n4}).
With the increasing $r$, the local maximum at $x=x_0=0$ emerges and the non-equilibrium stationary density becomes trimodal.
The local maximum induced at $x_0=0$ is produced by stochastic resetting.
More precisely, the resetting introduces a source of particles at $x_0$, which for sufficiently large $r$ builds a sustainable cusp (mode) at this point, see Figs.~\ref{fig:st-x4} --~\ref{fig:pd-x4} and Appendix~\ref{sec:res-app}.
At the same time, the heights of the outer peaks diminish, because with the increasing resetting rate the exploration of distant points is limited.
Consequently,  possibilities of creating  outer peaks are decreased and the central mode grows and accumulates increasing fraction of the probability mass.
Finally, for large enough $r$, the stationary state is unimodal, because exploration of the space is virtually fully eliminated.

Conclusions drawn from the examination of stationary states are presented in the form of the phase diagram, see Fig.~\ref{fig:pd-x4}, which shows the positions of modal values $\xm$ as a function of the resetting rate $r$.
From Fig.~\ref{fig:pd-x4} it is possible to read the number of modes and their locations.
Within the diagram, there are three distinct regions corresponding to the bimodal (low $r$ --- $r<0.02$), trimodal (intermediate $r$ --- $0.02 < r < 0.4$) and unimodal (large $r$ --- $r \geqslant 0.4$) stationary states.
The resetting induced mode at $x=x_0=0$ is visible for a large enough resetting rate, i.e, $r>0.02$.

\begin{figure}[!h]
	\centering
\includegraphics[angle=0,width=0.95\columnwidth]{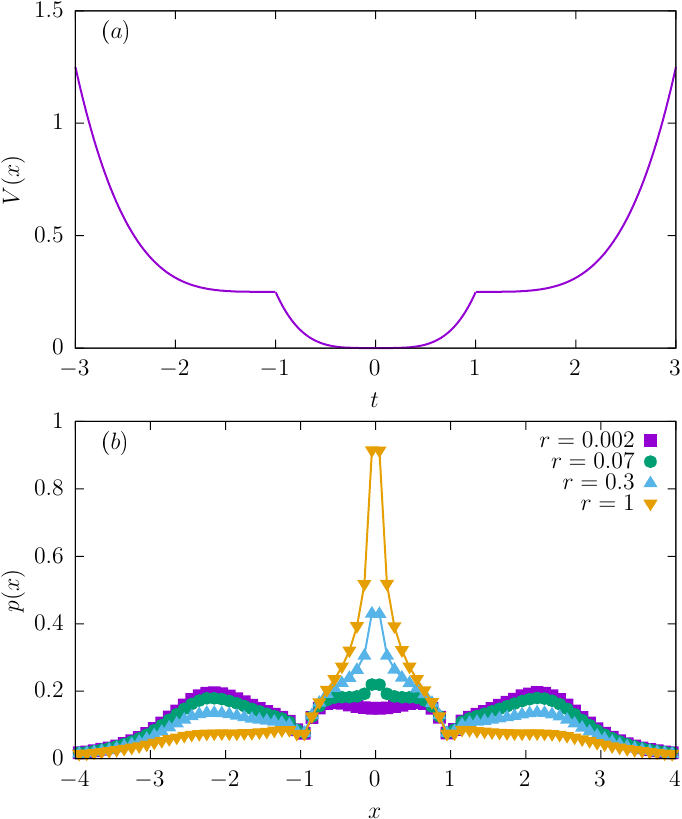}\\
   	  \caption{
   	  The same as in Fig.~\ref{fig:st-x4} for the ``glued'' potential, see Eq.~(\ref{eq:glued}).
   	  }
	\label{fig:st-glued}
\end{figure}

\begin{figure}[!h]
	\centering
\includegraphics[angle=0,width=0.95\columnwidth]{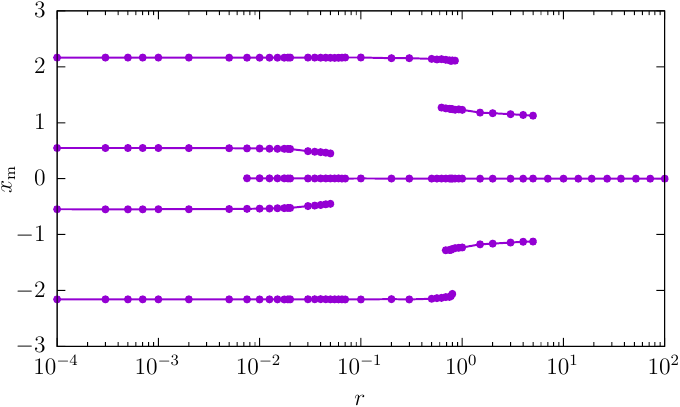}\\
   	  \caption{
   	  The same as in Fig.~\ref{fig:pd-x4} for the ``glued'' potential, see Eq.~(\ref{eq:glued}).
   	  }
	\label{fig:pd-glued}
\end{figure}

The quartic setup, which was studied so far, can be further generalized.
Closer inspection is needed for more general single-well potentials \cite{capala2019multimodal} as they, in the absence of resetting, can produce stationary states with the higher number of modes than two.
For instance we can use the ``glued'' potential
\begin{equation}
	V(x) = \left\{
	\begin{array}{lcl}
  	  \frac{(x+1)^4}{16} + \frac{1}{4}  & \mbox{for} & x<1  \\
   	  \frac{x^4}{4} &  \mbox{for} & |x| \leqslant 1 \\
   	  \frac{(x-1)^4}{16} + \frac{1}{4}  & \mbox{for} & x>1  \\
	\end{array}
	\right.,
	\label{eq:glued}
\end{equation}
depicted in Fig.~\ref{fig:st-glued}(a), which belongs to the class of single-well potentials capable of producing quadrimodal stationary states \cite{capala2019multimodal}.
The potential~(\ref{eq:glued}) is tuned in such a way that the internal quartic part allows for emergence of two modes within the $|x|<1$ domain, while outer parts produce two more modes.
Under Poissonian resetting to $x=0$ the resetting induced peak (mode) located at the origin emerges faster than outer peaks disappear, therefore the stationary states can have five modes, see Fig.~\ref{fig:st-glued}(b).
With the the increase in $r$, modes located at $|\xm|\approx 0.5$ disappear first, and lastly outer modes vanish as well.
Finally, Fig.~\ref{fig:pd-glued} shows the phase diagram for the ``glued''  potential.
From the phase diagram it is clearly visible that initial quadrimodal stationary density changes into unimodal density via five-, three-, four- and three-modal intermediate non-equilibrium stationary states.
This indicates that transition from quadrimodal to unimodal density for the ``glued'' potential is more complex than for a smooth potential.
First of all, with the increasing resetting rate the restarting induced mode at the origin emerges.
Next, further increase of $r$ strengthens central mode until the internal modes at $|\xm|\approx 0.5$ disappear.
After removal of internal modes the stationary states are trimodal.
Subsequent growth in the resetting rate further weakens the heights of external modes and creates modes at $|x|\approx 1$.
Since external modes ($x \approx \pm 2$) are produced by long jumps (jumps ending at $|x| \gg 2$), there is an intermediate range ($0.6 < r < 0.9$) when (weak) modes at gluing points emerge but (weak) external modes are still visible as resetting is unable to fully eliminate consequences of long jumps.
For larger $r$ ($0.9<r<4$) effects of long jumps are efficiently eliminated because the motion is restarted prior to emergence of modes at $x \approx \pm 2$,  while intermediate jumps still result in accumulation of the probability mass around $x\approx \pm 1$ which is not fully depleted by resetting.
Finally, for a large enough resetting rate the non-equilibrium stationary states attain the unimodal shape.

%
%
\section{Summary and conclusions \label{sec:summary}}

\begin{figure}[!h]
	\centering
\includegraphics[angle=0,width=0.95\columnwidth]{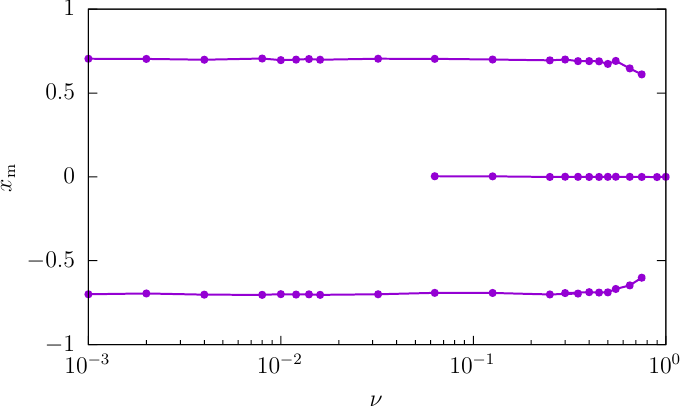}\\
   	  \caption{
   	  The same as in Fig.~\ref{fig:pd-x4} for the Pareto distribution of interresetting times with $\tau_0 = 0.1$, see Eq.~(\ref{eq:pareto}).
   	  }
	\label{fig:pd-x4-pareto}
\end{figure}
The results reported here confirm that the (Poissonian) stochastic resetting to a single, fixed point can be used to control the modality of non-equilibrium stationary states.
The consequences of restarting are especially well visible for the Cauchy quartic oscillator for which resetting to a fixed point (origin) can change the modality of the stationary state.
With the increasing resetting rate, the initial bimodal stationary state attains a trimodal form and finally becomes unimodal.
The richer scenario is recorded for the ``glued'' potential for which quadrimodal stationary state can be changed into five-, three- or four-modal state.
Alternatively, instead of resetting to a fixed point, it is possible to restart a motion from randomly selected points.
In such a situation, the stationary state can be of any multimodality, because resetting events are capable of producing peaks at each restarting point.
Nevertheless, such a mechanism does not fully utilize abilities of L\'evy flights to explore space, because it relies on the fact that under stochastic resetting the stationary density is the composition of stationary states corresponding to resetting to fixed points.
Stochastic resetting alone introduces sources of particles at the resetting positions which can produce modes at these points.

The Poissonian (constant rate) resetting is one of possible (temporal) resetting scenarios, i.e., times of starting over are independent of state of the system, e.g., position.
In general, stochastic resetting assumes that the motion is started anew at random times.
Therefore, it is also possible to study other types of stochastic restarting like sharp resetting  or power-law resetting
\begin{equation}
p(\tau)=
\frac{\nu \tau_0^\nu}{\tau^{\nu+1}},
\label{eq:pareto}
\end{equation}
where $\tau_0$ ($\tau_0>0$) is a scale parameter, while $\nu$ ($\nu>0$) is the shape parameter (tail index).
For the Pareto distribution, see Eq.~(\ref{eq:pareto}), the mean value exists for $\nu>1$, while the variance is finite for $\nu>2$.
Therefore, for $\nu\leqslant 1$ the mean interresseting time diverges. It becomes finite for $\nu>1$, but in the region $1<\nu \leqslant 2$ it is still characterized by the diverging variance.
In our setup, the particle motion is bounded by the external potential producing the (non-equilibrium) stationary state already in the absence of restarting.
Therefore, the stochastic resetting at power-law times does not affect the existence of stationary states but it can influence only their shape.
For the motion in $V(x)=x^4/4$ with the power-law distribution of interresseting times, see Eq.~(\ref{eq:pareto}),  with respect to the modality of non-equilibrium stationary states, qualitatively the same behavior (although at different values of parameters, c.f. Fig.~\ref{fig:pd-x4-pareto}) like for the Poissonian resetting is observed.
Interestingly, due to the fact that for uninterrupted motions in quartic potentials, non-equilibrium stationary states exist, the change in modality is not associated with the existence of the mean interresetting time like for the free motion\cite{nagar2016diffusion} with resets at power-law times.
Nevertheless, detailed exploration of properties of non-equilibrium stationary states under power-law resetting calls for further studies.
Finally, sharp resetting  periodically restarts evolution of the probability distribution making it time dependent. More precisely, at resets the whole probability mass is relocated to the restarting position. Afterwards it widens till the next start over.

The quartic and the ``glued'' potentials are just two exemplary potentials which can be used to demonstrate emergence of additional modes because of stochastic resetting.
It is still possible to use other types of potentials\cite{capala2019multimodal}.
However, the most important limitation of such studies arises due to the difficulty in estimating the exact values of the resetting rate corresponding to various numbers of modes, as the number of modes is obtained from the histogram.
Moreover, the most interesting potentials are of such a type that in the absence of resetting the stationary density has minimum at the origin, because starting anew from the origin can induce emergence of the mode at this point.

%
%
\section*{Acknowledgements}

We gratefully acknowledge Poland’s high-performance computing infrastructure PLGrid (HPC Centers: ACK Cyfronet AGH) for providing computer facilities and support within computational grant no. PLG/2023/016175.
The research for this publication has been supported by a grant from the Priority Research Area DigiWorld under the Strategic Programme Excellence Initiative at Jagiellonian University.

\section*{Data availability}
The data  (generated randomly using the model presented in the paper) that support the findings of this study are available from the corresponding author (PP) upon reasonable request.

%
%
\appendix

\section{Stochastic resetting\label{sec:res-app}}
For completeness of the presentation, we repeat the basic information regarding stochastic resetting \cite{evans2011diffusion-jpa,evans2020stochastic}.
Under the action of stochastic Poissonian resetting and the Gaussian white noise, for a particle moving in the potential $V(x)$, the probability density $p(x,t)=p(x,t|x_0,0)$ evolves according to
\begin{eqnarray}
	\label{eq:fpe-reset-point}
	\frac{\partial p(x,t)}{\partial t} & = &  \sigma^2 \frac{\partial^2 p(x,t)}{\partial x^2}  + \frac{\partial}{\partial x}\left[ V'(x)p(x,t) \right]\\
	& - &   r p(x,t) + r \delta(x-x_0) . \nonumber
\end{eqnarray}
In the above equation, $-r p(x,t)$ describes a sink while the $r\delta(x-x_0)$ term represents the source of probability  at $x_0$, which could build a mode at the resetting position $x_0$.
The stationary solution $p(x)$ of Eq.~(\ref{eq:fpe-reset-point}) fulfills
\begin{equation}
	\sigma^2 \frac{\partial^2 p(x)}{\partial x^2}  + \frac{\partial}{\partial x}\left[ V'(x)p(x) \right] - r p(x) = - r \delta(x-x_0).
	\label{eq:fpe-reset}
\end{equation}
For some simple setups, $p(x)$ can be found analytically \cite{evans2011diffusion-jpa,evans2020stochastic,stanislavsky2021optimal}.

If instead of resetting to a fixed position $x_0$ the reset is performed to a set of points distributed according to $\mathcal{P}(x)$, Eq.~(\ref{eq:fpe-reset-point}) attains the following form \cite{evans2011diffusion-jpa,evans2020stochastic}
\begin{eqnarray}
\label{eq:fpe-reset-set}
	\frac{\partial p(x,t)}{\partial t} & = &  \sigma^2 \frac{\partial^2 p(x,t|x_0)}{\partial x^2}  + \frac{\partial}{\partial x}\left[ V'(x)p(x,t) \right] \\
	& - &  r p(x,t) + r \mathcal{P}(x) \nonumber
\end{eqnarray}
The stationary state $p(x)$ for Eq.~(\ref{eq:fpe-reset-set}) is the composition of $p(x|z)$
\begin{equation}
	p(x)=\int \mathcal{P}(z) p(x|z)dz,
	\label{eq:superposition}
\end{equation}
where $p(x|z)$ is the stationary solution for resetting to $z$.
Consequently, for $\mathcal{P}(x)$ being the normalized sum of Dirac's delta, under stochastic resetting the stationary state is the sum of stationary states for each $x_0$, see Ref.~\onlinecite{pal2015diffusion}.
Consequently, such a solution could be multimodal with modes located at the restarting positions $x_0$.
Eqs.~(\ref{eq:fpe-reset-point}) -- (\ref{eq:fpe-reset-set}) can be extended to describe systems driven by $\alpha$-stable noises.
In such a case, Eqs.~(\ref{eq:fpe-reset-point}) -- (\ref{eq:fpe-reset-set}) becomes of the fractional order, see Eq.~(\ref{eq:ffpe}) and Refs.~\onlinecite{podlubny1999,kilbas2006}, but importantly Eq.~(\ref{eq:superposition}) stays intact.

%
%

\section*{References}

\def\url#1{}

\end{document}